\documentstyle[epsfig,aps,prb,twocolumn,floats,eqsecnum]{revtex}

\begin{document}

\newcommand{\gsim}{
\,\raisebox{0.35ex}{$>$}
\hspace{-1.7ex}\raisebox{-0.65ex}{$\sim$}\,
}

\newcommand{\lsim}{
\,\raisebox{0.35ex}{$<$}
\hspace{-1.7ex}\raisebox{-0.65ex}{$\sim$}\,
}

\newcommand{\const}{ {\rm const} }
\newcommand{\arctanh}{ {\rm arctanh} }

\bibliographystyle{prsty}

\title{ \begin{flushleft}
{\small \em published in}\\
{\small 
PHYSICAL REVIEW LETTERS 
\hfill
VOLUME 80
%{\normalsize 57}, 
NUMBER 13
%{\normalsize 17}
\hfill 
30 MARCH 1998
%{\normalsize 1}
%MAY
%{\normalsize 1996}-1 
%{\normalsize 3250-3256}
}\\
\end{flushleft}  
Pyrochlore Antiferromagnet: A Three-Dimensional Quantum Spin Liquid 
}

\author{
Benjamin~Canals \cite{e-can} and Claudine Lacroix \cite{e-lac}
}

\address{
Laboratoire de Magn\'etisme Louis N\'eel, CNRS, B.P. 166, 38042 Grenoble Cedex 9, France\\
%}
%\date{18 November 1997}
%\maketitle
%\abstract{
\smallskip
%{\rm (\today) }
{\rm (Received 12 December 1997) }
\bigskip\\
\parbox{14.2cm}
{\rm
The quantum pyrochlore antiferromagnet is studied by perturbative expansions and exact diagonalization of small clusters. We find that the ground state is a spin-liquid state: The spin-spin correlation functions decay exponentially with distance and the correlation length never exceeds the interatomic distance. The calculated magnetic neutron diffraction cross section is in very good agreement with experiments performed on Y(Sc)Mn$_2$. The low energy excitations are singlet-singlet ones, with a finite spin gap.
\smallskip
\begin{flushleft}
PACS number(s): 75.10.Jm, 75.50.Ee, 75.40.-s
\end{flushleft}
} 
} 
\maketitle

Since Anderson proposed the resonant valence bond (RVB) wave function for the triangular lattice \cite{and72}, there has been a lot of attention focused on frustrated lattices, because they might have a spin-liquid-like ground state in two- or three-dimensional lattices. 
The frustrated systems can be classified into two different categories: structurally disordered systems and periodic lattices. 
Among the last one the ground state of the $S=\frac{1}{2}$ quantum Heisenberg antiferromagnet on the kagom\'e and pyrochlore lattices is expected to be a quantum spin liquid (QSL). 
The common point of these two lattices is the high degree of frustration as they both belong to the class of the ``fully frustrated lattices.'' 
The classical mean field description indicates a pathological spectrum with an infinite number of zero energy modes which prevents any magnetic phase transition and produces an extensive zero temperature entropy \cite{chu92,cha93,har92}. 
The classical critical properties are nonuniversal for both systems \cite{rei921}.

The pyrochlore lattice consists of a 3D arrangement of corner sharing tetrahedra (Fig. 1). 
All compounds which crystallize in the pyrochlore structure exhibit unusual magnetic properties: 
Two of them, FeF$_3$ and NH$_4$Fe$^{2+}$Fe$^{3+}$F$_6$ \cite{fer86}, are known to have a noncollinear long range ordered magnetic structure at low temperature; 
the other compounds do not undergo any phase transition, but many of them behave as spin glasses, although there is no structural disorder at all. 
It is remarkable that frustration in a periodic lattice may give rise to ageing and irreversibility so that a conpound such as Y$_2$Mo$_2$O$_7$ has been considered as a ``topological spin glass'' \cite{cha93,gin97}.

The problem of ordering in the pyrochlore lattice was initiated by Anderson \cite{and56} who predicted that only long range interactions are able to stabilize a N\'eel-like ground state. 
More recently, mean field studies \cite{rei91} have confirmed these predictions; from classical Monte Carlo calculations \cite{rei921,lie86,rei922} it was concluded that this system does not order down to zero temperature, but any constraint will induce magnetic ordering \cite{bra94}.

The only attempt to describe the quantum $S=\frac{1}{2}$ Heisenberg antiferromagnet on the pyrochlore lattice has been done by Harris {\it et al.} \cite{har91} who studied the stability of a dimer-type order parameter and showed that quantum fluctuations play a crucial role. 
In this Letter, we show that the ground state exhibits a QSL behavior: 
By applying a perturbative approach to the density operator we show that spin-spin correlations decay exponentially with distance at all temperatures with a correlation length that never exceeds the interatomic distance. 
Fluctuations select collinear modes, but the amplitude of these modes is extremely small. 
Exact diagonalization of small clusters shows that the spectrum of the pyrochlore lattice looks like the kagom\'e one, with a singlet-triplet spin gap and no gap for the singlet-singlet excitations. 
We also use our developpement to calculate the neutron magnetic cross section. 
The results are in very good agreement with previous experimental results on Y(Sc)Mn$_2$ \cite{bal96}.
\begin{figure}[t]
\unitlength1cm
\begin{picture}(5,6)
\centerline{\epsfig{file=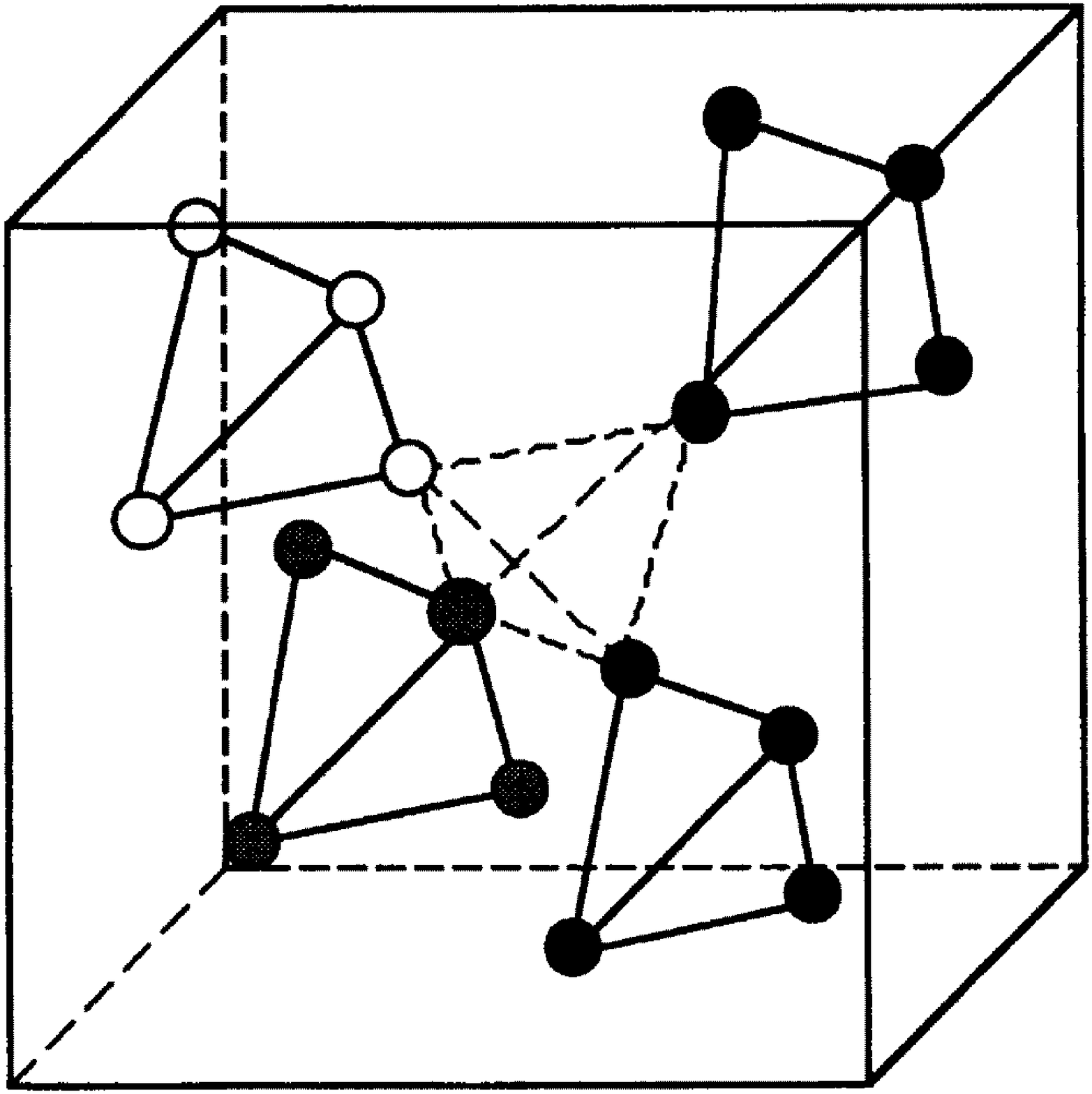,angle=0,width=7cm}}
\end{picture}
\\
\caption{
Description of the pyrochlore lattice as an fcc lattice of tetrahedra. Solid lines connect sites interacting with exchange $J$, dashed lines with exchange $J^{\prime}$.
}
\end{figure}
{\it Model}.--The Hamiltonian of the quantum Heisenberg model on the pyrochlore lattice is

\begin{equation}
H = - J \sum_{\langle i,j \rangle} {S_i . S_j},
\end{equation}

where the summation is taken over the nearest neighbors (NN) sites. 
$J$ is the negative exchange coupling. 
We describe the pyrochlore lattice as a fcc Bravais lattice with a tetrahedral unit cell. 
Each cell is exactly diagonalized and coupling between cells is taken into account perturbatively: 
We call $J^{\prime}$ the exchange interaction between NN sites in different tetrahedra (Fig. 1) and we make an expansion in powers of $\lambda = J^{\prime} / J$. 
The Hamiltonian is rewritten as $H=H_{0}(J)+H_{1}(J^{\prime})$, where $H_0$ describes the NN interactions within a tetrahedron and $H_1$ the NN intertetrahedra interactions. 
By using this trick we have transformed the initial $S=\frac{1}{2}$ pyrochlore lattice into a fcc lattice with 16 states per ``site'' (four spins $\frac{1}{2}$). 
All of the thermodynamical quantities can be obtained from the density operator, which we calculate in perturbation with respect to $\lambda$:
\begin{equation}
\rho = e^{-\beta H} = \rho_0 + ... + \rho_n + \Theta (\lambda^n),
\end{equation}
where
\begin{eqnarray}
\rho_n = & (-1)^n & \int_{0}^{\beta}
                    \int_{0}^{\beta_1} ...
                    \int_{0}^{\beta_{n-1}}
                    d\beta_1 ... d\beta_n e^{-(\beta - \beta_1) H_0}  \nonumber\
 \\
         & \times & H_1 ... e^{-(\beta_{n-1} - \beta_n) H_0} H_1 e^{-\beta_n H_0}.
\end{eqnarray}

In order to evaluate the different terms of the development, we have derived a diagrammatic method which allows a systematic expansion. 
This method will reported elsewhere. 
All of the following quantities have been evaluated {\it analytically} to the second order in $\lambda$ by implementing a formal program in Mathematica on a Silicon Graphics. 
For an indication, the evaluation of the spin-spin correlation functions from the first to the sixteenth neighbors took $1 \frac{1}{2}$ months of CPU time. 
The method was tested on the spin-$\frac{1}{2}$ chain for which the unit cell was two NN sites and the development was made to fourth order in $\lambda$. 
Using the analytical density operator and fixing $\lambda = 1$, we obtained quite satisfactory results, the difference with those of Bonner and Fisher \cite{bon64} being less then 2 \% at low temperature. 
Furthermore, our method is more efficient when the characteristic length of the ground state is short, as expected in a spin liquid. 
Thus, the deduced characteristic length will be a control parameter of our calculations. 
Moreover, contrary to usual high temperature expansion, our method yields to a
nondivergent expansion when $T \rightarrow 0 $.

{\it Spin-spin correlations.}--Let us consider a reference site $S_0$ on the lattice. 
We define $\langle S_{0}.S_{d} \rangle = C_d$ as the correlation function between this site and a site at a distance $d$ in the lattice. 
This function is easily evaluated as
\begin{equation}
C_d = \frac{1}{Z} Sp[S_0 . S_d \exp(- \beta H)],
\end{equation}
where $Z$ is the partition function. 
The development was made up to second order in $\lambda$, which allows us to calculate $C_d$ up to the $16th$ neighbors. 
At any temperature between $T = 10 |J|$ and $T = |J|/100$, we find that these correlations oscillate in sign and decay exponentially with the distance [Fig. 2(a)]. 
From these curves, we can extract a correlation length $\xi$ defined as $|C_d| \propto exp(-d/\xi)$. 
This length is temperature dependent and never exceeds one interatomic distance down to zero temperature [Fig. 2(b)]; 
the broad feature at $T \approx J$ shows the limitations of the method in this energy scale. 
\begin{figure}[t]
\unitlength1cm
\begin{picture}(11,10)
\centerline{\epsfig{file=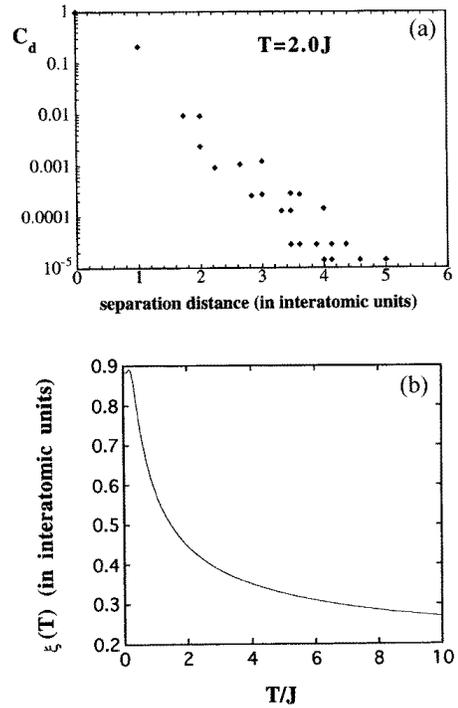,angle=0,width=6cm}}
\end{picture}
\caption{
(a) Absolute value of the spin-spin correlatin functions as a function of distance at $T = 2|J|$. (b) Correlation length deduced from the analysis of the (a)-type graphs at all temperatures.
}
\end{figure}
Assuming that $\xi$ could be underevaluated for $T \leq J$ by our method, we
extrapolated $\xi (T)$ from the $T \geq J$ regime to the low $T$
regime. 
Whatever the law we used (algebraic, exponential, stretched
exponential), we found a saturation of the correlation length around the
interatomic distance. 
Thus, our calculations are self-consistently controlled
as the value of $\xi$ at $T = 0K$ is much smaller than the real spatial
extension of our developpement (five interatomic distances).

The total spin can be obtained from the spin-spin correlation functions as
\begin{equation}
\langle ( \sum_{i=1}^{N} S_i )^2 \rangle = NS(S+1) + 6NC_1 + 12NC_2 + ...
\end{equation}
Since we have shown that these correlation functions are rapidly decreasing with distance, we can retain only a few terms in this expansion. 
Cutting the summation to the second neighbors, we verify that the ground state is a singlet:
\begin{equation}
\frac{1}{N} \langle ( \sum_{i=1}^{N} S_i )^2 \rangle (T=0) \cong 0.02
\end{equation}
This result shows that our method is satisfactory even at a very low temperature, essentially because the correlations inside each unit cell are calculated exactly, and in a QSL, only short range correlations are important.

{\it Static structure factor}.--For classical spins, the spectrum of the pyrochlore lattice is very peculiar \cite{rei91}: 
It possesses two branches of zero energy modes. 
In order to study the effect of quantum and thermal fluctuations on these modes, we evaluated the structure factor,
\begin{equation}
S^{m,n} (q) = \sum_{d} {C_d e^{iq.R_{d}^{m,n}}}
\end{equation}
where $m$ and $n$ are the indices of a site in a tetrahedral unit cell [$(m,n)
\in \{1,2,3,4\}^2$]. 
$C_d$ is defined in Eq. (0.4), $q$ is a vector of the
first Brillouin one (BZ), and $R_{d}^{m,n}$ is the vector of length $d$ that links
the sites of type $m$ and $n$. 
For
each $q$ this structure factor is a $4 \times 4$ matrix whose eigenvalues
give the fluctuation modes of the system, the lowest energy mode corresponding
to the largest eigenvalue $\omega_M (q)$. 
To the first order in $\lambda$,
$\omega_M (q)$ remains nondispersive over the entire BZ, and degeneracy is not
lifted. 
To the second order, a maximum appears on the axis $\Delta$ of the
BZ (Fig. 3). 
This maximum corresponds to a collinear phase where
the total spin vanishes on each tetrahedron (see the inset of Fig. 3), and the phase between two
neighbouring tetrahedra is equal to $\pi$. 
This confirms the results obtained for classical spins by previous authors \cite{rei922,hen87}. 
We note that the degeneracy is very weakly lifted ($1/10^6$ of the width of the spectrum), but this is not a numerical artifact since, in our method, the precision can be as small as we want. 
A similar behavior was also observed in the kagom\'e lattice \cite{els94}.
\begin{figure}[t]
\unitlength1cm
\begin{picture}(11,7)
\centerline{\epsfig{file=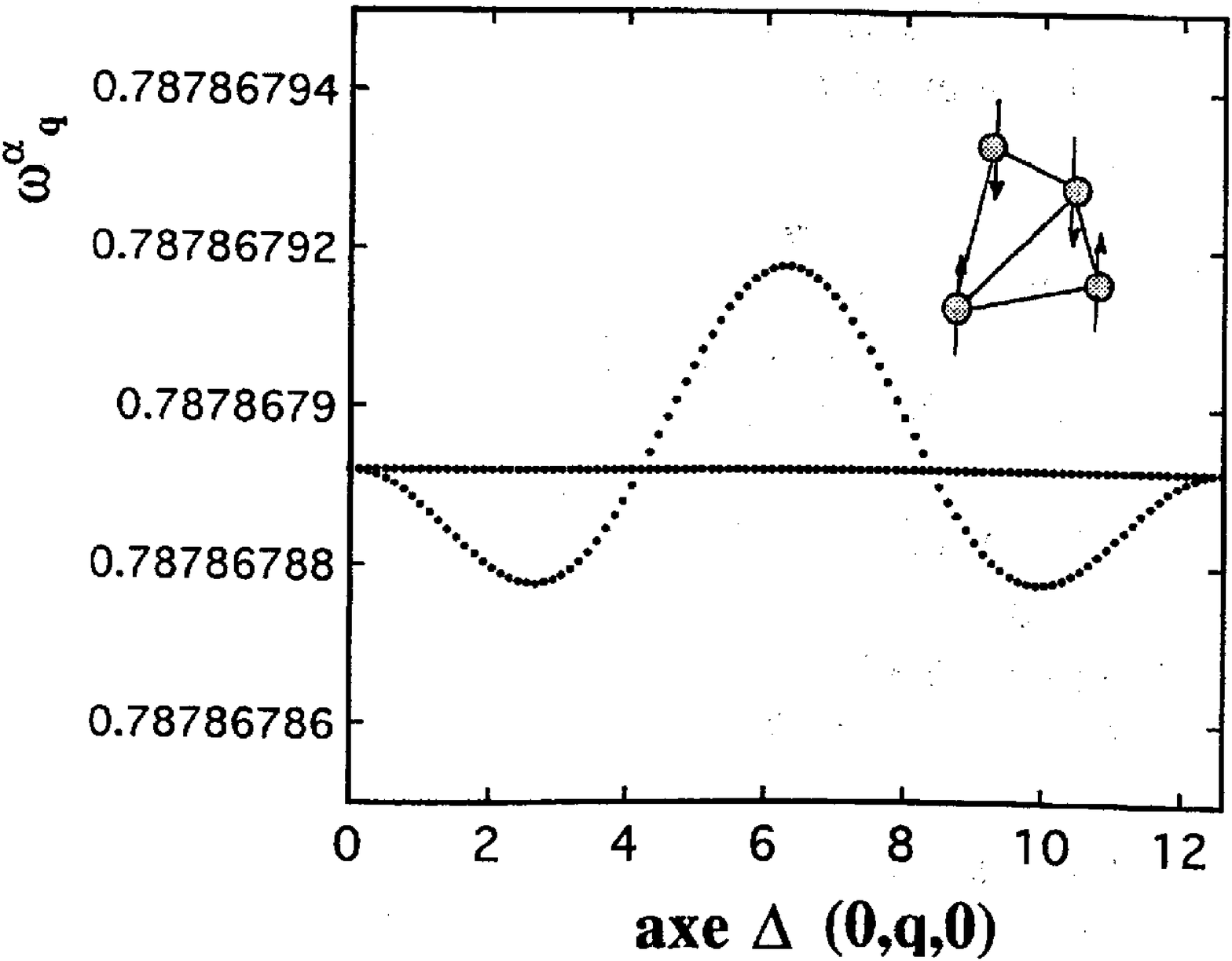,angle=0,width=8cm}}
\end{picture}
\caption{
The two largest eigenvalues of the structure factor along the $\Delta$ axis. The maximum corresponds to the collinear phase shown in the inset.
}
\end{figure}
{\it Neutron cross section}.--The neutron magnetic cross section can be expressed in terms of the correlation functions as
\begin{equation}
\frac{d^2 \sigma}{d \Omega d \omega} =
\sum_{m,n}^{\{1,2,3,4\}^2} {e^{-i \kappa .(T_m - T_n)} U_{m,n} (Q,\omega)},
\end{equation}
where
\begin{eqnarray}
U_{m,n} (Q,\omega)  = & & \sum_{i,j} {e^{-q.(R_i - R_j)} e^{-q.(T_m - T_n)}} \\
& \times & \int_{- \infty}^{+ \infty}
\langle S_{i,m} (0)S_{j,n} (t) \rangle e^{-i \omega t} dt.
\end{eqnarray}
$T_m$ and $R_i$ are the translations that define the position of a site of
type $m$ in the unit cell i of the space group $Fd \mbox{\={3}} m$; 
$Q= \kappa +
q$, where $\kappa$ is a vector of the reciprocal lattice and $q$ belongs to
the first BZ. 
From the static correlation functions calculated above, we
obtain the total magnetic cross section $d \sigma / d \Omega$. 
We report the results as a contour plot in the reciprocal ([$00h$],[$hh0$]) plane (Fig. 4). 
The absence of a signal in the first BZ indicates that the ground state is a singlet. 
The intensity is maximum at about $Q_1 = [200] \pm [\frac{3}{4} \frac{3}{4} 0]$ or $Q_{1}^{\prime} = [002] \pm
[\frac{3}{4} \frac{3}{4} 0]$ (Fig. 4). 
These maxima are associated with long range correlations describing a structure where consecutive tetrahedra are in phase. 
This is different from the result obtained above from the structure factor. 
In fact, we have also calculated the cross section in the ([$h00$],[$0h0$]) plane and found another maximum at $Q_2 = [210]$ which corresponds to a phase
$\pi$ between tetrahedra as expected. 
Comparison of the intensities of the peaks at $Q_1$ and $Q_2$ favors the second structure, but the difference is small. 
So we conclude that there are two characteristic modes in this system, a $\pi$-dephased one and an in-phase one, with a larger weight for the first mode.
\begin{figure}[t]
\unitlength1cm
\begin{picture}(11,7)
\centerline{\epsfig{file=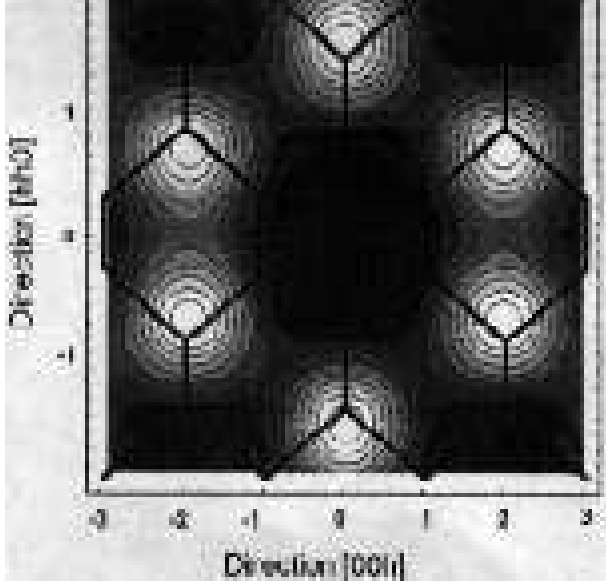,angle=0,width=8cm}}
\end{picture}
\\
\caption{
Map of the neutron cross section in the plane ([$00h$],[$hh0$]). The white regions indicate the maximum of intensity in the {\bf q} space.
}
\end{figure}
It is interesting to compare our results (Fig. 4) with the measurements made on Y(Sc)Mn$_2$ by Ballou {\it et al.} \cite{bal96}: The experimental results correspond exactly to our result (Fig. 4). 
In fact, the measurements were made for a given energy $\omega$, but the shape of the neutron cross section in reciprocal space was found to be nearly independent of the energy \cite{bal96}, as if $ \chi (q,\omega ) = f(q)g( \omega )$. 
Thus, integrating the experimental
results over the energy does not change the q dependance and we can make a
direct comparison between our calculations and the measured map. 
First we
reproduce all the main features previously obtained, i.e the absence of a
signal in the firt BZ associated with the singlet ground state and the maxima at $Q_1$ and $Q_{1}^{\prime}$. 
Second, we find a $\pi$-dephased mode in the ([$h00$],[$0h0$]) plane that was also observed in the same compound. 
The existence of these two modes could explain the first order character of the transition observed in the pure YMn$_2$ compound \cite{shi88,shi94} associated with the freezing of the mode mixture in the $\pi$-dephased one. 
Finally, the half-width of each peak provides information on the correlation length: 
It is found to be of one interatomic distance in experiments as well as in our calculations. 
All these results indicate that a large part of the physics of Y(Sc)Mn$_2$ is related to quantum fluctuations in this highly frustrated structure, although the magnetism of Mn is not localized but itinerant.

{\it Low energy part of the spectrum}.--We performed exact diagonalization of small clusters, up to 12 spins $\frac{1}{2}$. 
These sizes are not large enough to give quantitative conclusions, but allow a first characterization of the spectrum. 
We find that the ground state is a singlet with a spin gap between this singlet and the lowest triplet ($\Delta \approx 0.7 |J|$). 
Inserted in this gap, there are several singlet states whose number is growing with the cluster size. 
This indicates that the lowest energy excitations are singlet-singlet-like. 
A similar property was also observed in exact diagonalizations performed on the kagom\'e lattice \cite{lec97}. 
It seems that both systems belon to the same class of QSL while the 1D integer spin chains, for which the lowest energy excitations are spin excitations (singlet-triplet), are different. 
In our case, the relevant parameter is the topological frustration, while in 1D Heisenberg spin liquids the low dimensionality plays a crucial role.

In conclusion, we have studied the quantum Heisenberg spin-$\frac{1}{2}$ Hamiltonian on the pyrochlore lattice. 
It appears from the spin-spin correlations and the low energy spectrum that the ground state of this system is a QSL similar to the kagom\'e lattice ground state. 
Using ou results we calculated the magnetic neutron cross section, and reproduced almost exactly the map experimentally observed in Y(Sc)Mn$_2$. 
This confirms that this compound is a 3D QSL.

The pyrochlore antiferromagnet is certainly the first example of a 3D QSL. 
For this class of QSL the dimensionality of the lattice seems to play a minor role, but we can expect that any perturbation will deeply modify the low energy spectrum. 
Thus, such a disordered magnetic ground state should be extremely sensitive to chemical disorder. In fact, in several compounds, such as Al-doped Y(Sc)Mn$_2$ \cite{shi94}, Y$_2$Mo$_2$O$_7$ \cite{mek95}, and Al-doped $\beta$-Mn \cite{nak97}, disorder, even nondetectable, transforms a QSL into a quantum spin glass state. 
All of these compounds have in common unconventional behaviors (i.e, the susceptibility is spin-glass-like, while the low T specific heat increases as $T^2$ and strong fluctuations of the local magnetization are observed in the ``frozen'' state). 
A quantitative understanding of disorder effects in QSL requires more studies, both experimental and theoretical.

We are grateful to R. Ballou, C. Lhuillier, and F. Mila for helpful discussions and comments.\\

\end{document}